\documentclass[pra,twocolumn,showpacs,floatfix,titlepage,superscriptaddress,here]{revtex4-1}
\usepackage{amsfonts}
\usepackage{amsmath}
\usepackage{amssymb}
\usepackage{graphicx}
\usepackage{comment}
\usepackage{bm}
\usepackage{mathrsfs}
\usepackage{subfigure}
\usepackage{color}

\renewcommand{\phi}{\varphi}
\renewcommand{\>}{\rangle}
\newcommand{\<}{\langle}

\newcommand{\ket}[1]{|#1\>}
\newcommand{\bra}[1]{\<#1|}

\newcommand{\be}{\begin{equation}}
\newcommand{\ee}{\end{equation}}
\newcommand{\bea}{\begin{eqnarray}}
\newcommand{\eea}{\end{eqnarray}}

\newcommand{\Comp}{\mathbb{C}}

\newcommand{\Real}{\mathbb{R}}
\newcommand{\aver}[1]{\left \langle #1\right \rangle}
\newcommand{\hlf}{\mbox{$\frac{1}{2}$}}
\newcommand{\hs}{\hat{\sigma}}
\newcommand{\ha}{\hat{a}}

\begin{document}
\title{Statistics of a quantum-state-transfer Hamiltonian in the presence of disorder}
\author{Georgios~M.~Nikolopoulos}
\affiliation{Institute of Electronic Structure \& Laser, FORTH, P.O.Box 1385, GR-71110 Heraklion, Greece}

\date{\today}

\begin{abstract}
We present a statistical analysis on the performance of a protocol for the faithful transfer of a quantum state in finite qubit or spin chains, in the presence of diagonal and off-diagonal disorder. It is shown that the average-state fidelity, typically employed in the literature for the quantification of the transfer, may overestimate considerably the performance of the protocol in a single realization, leading to faulty conclusions about the success of the transfer. 
\end{abstract}

\maketitle

\section{Introduction}

The faithful transfer of quantum states between two or more spatially separated qubits of a quantum network has attracted considerable interest over the last decade \cite{QSTRevs}. 
In its simplest form, the problem of quantum state transfer (QST) pertains to the 
quest for Hamiltonians (protocols) that ensure the transfer of a state between the two ends of a qubit chain at a prescribed time \cite{design}.  
The qubit chain may be represented by an actual spin chain in a liquid or solid-state NMR system \cite{liquidNMR,solidNMR}, nitrogen vacancies in diamond \cite{NVqst},  or by other realizations pertaining to optical lattices \cite{BBVB11}, coupled quantum dots \cite{weNPL,YBB10}, superconducting qubits \cite{Romito,Strauch}, or waveguides \cite{BNT12}.  The first proof-of-principle experiment on a QST 
Hamiltonian with engineered couplings has been presented recently in the framework of waveguides  \cite{BNT12}. 

Disorder is expected to be present in any physical realization of  qubit (spin) chains, irrespective of the experimental platform. Thus, theoretical investigations on the robustness of QST Hamiltonians against disorder is an essential 
step toward any experimental realization, as well as for the classification of different protocols, with respect to their 
performance under realistic conditions. To the best of our knowledge, so far such investigations have relied on an input-independent  fidelity obtained by averaging over all possible input states \cite{YBB10,Romito,Strauch,DeChiara,dapPRA10,Zwick,AlLin09}. This average-state quantity is evaluated for a large number of independent realizations (pertaining to randomly chosen disorder for the diagonal and off-diagonal elements of the Hamiltonian), and all the analysis of the performance of a protocol is based on the ensemble averaged input-independent  fidelity. 

In order for a QST Hamiltonian to be reliable and useful, however, its performance for a certain level of disorder has to be within  the acceptable levels for every single realization, irrespective of the input state. The disorder as well as the qubit state to be transferred may vary from realization to realization, but the efficiency and the success of the protocol under consideration have to be guaranteed.   Our aim in this work is to investigate whether and to what extent  the average-state  input-independent fidelity that has been used extensively in the literature, is capable of describing the performance of a Hamiltonian in a single realization. To this end, we consider one of the most studied QST Hamiltonians \cite{QSTRevs,weNPL,BNT12,Ekert04,KE05,GMN08,CVR11}, the robustness of which in the presence of disorder has been also investigated in various contexts \cite{weNPL,Strauch,DeChiara,dapPRA10,Zwick}. 

In Sec. \ref{sec2} we outline our model, whereas Secs. \ref{sec3} and \ref{sec4} are devoted to a thorough statistical analysis on the performance of the protocol in a single realization. It is shown that for a large class of input states the ensemble averaged input-independent fidelity overestimates the performance of the protocol, and may lead to faulty  conclusions  with respect to the success (or failure) of the transfer. Alternative forms of the fidelity turn out to be more reliable measures for the quantification of QST.  Our conclusions are summarized in Sec. \ref{sec5}. 

\begin{figure*}
\includegraphics[scale=0.5]{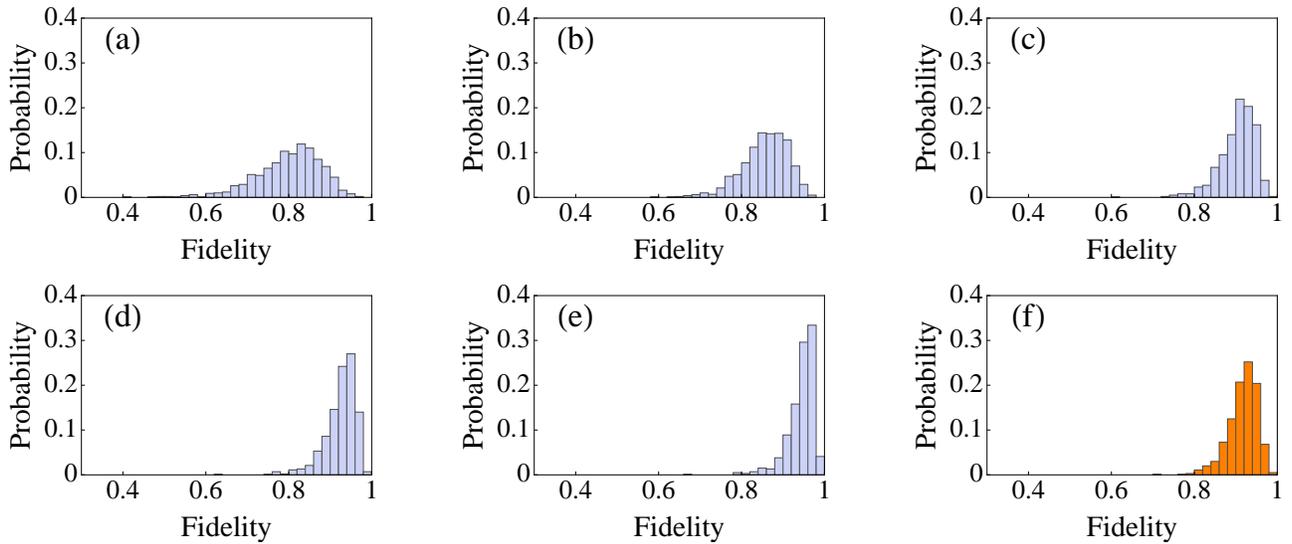}
\caption{(Color online) Probability distribution of the fidelity for $N=12$ and 1000 independent realizations. The histograms (a-e) correspond to $F_\psi(|\beta|^2,p,\Delta\varphi)$ for $|\beta|^2=1.0,\,0.8,\,0.6,\,0.5,$ and $0.4$, respectively. The histogram (f) corresponds to $\bar{F}(p,\Delta\varphi)$. Other parameters: $\sigma_\eta=\sigma_\xi=0.1$.}
\label{distro:fig}
\end{figure*}

\section{Model}
\label{sec2}

Our analysis pertains to QST spin-chain Hamiltonians of the form  
\be
\hat{\mathscr H} = -\hlf \sum_{k=1}^{N} \varepsilon_k \hs_k^z 
+ \hlf \sum_{k=1}^{N-1} J_k ( \hs_k^x \hs_{k+1}^x + \hs_k^y \hs_{k+1}^y), \label{HamSpCh} 
\ee
where $N$ the length of the spin chain, $\hs_k^{x,y,z}$ are the Pauli spin operators for the $k$th spin,
$\varepsilon_k$ is the energy separation between the spin-up and 
spin-down states playing the role of the local ``magnetic field'', 
and $J_k$ is the time-independent nearest-neighbor spin-spin interaction. 
This is a spin-chain Hamiltonian of the $XX$ type, which is isomorphic to the 
Hubbard Hamiltonian for {\em non-interacting} spinless fermions or hard-core bosons \cite{Sachdev}
\be
\hat{\mathscr H}  = -\sum_{k=1}^{N} \varepsilon_k \ha^{\dag}_k \ha_k 
+ \sum_{k=1}^{N-1} J_k ( \ha^{\dag}_k \ha_{k+1} + \ha^{\dag}_{k+1} \ha_k), 
\label{HamHub} 
\ee
where $\ha^{\dag}_k$ ($\ha_k$) is the particle creation (annihilation) 
operator at $k$th site with energy $\varepsilon_k$, and $J_k$ now plays the role
of tunnel coupling between adjacent sites $k$ and  $k+1$. 
From now on, the qubit basis states for a single spin (particle) are denoted by $\{\ket{0}, \ket{1}\}$. 

Our task is the faithful --- ideally perfect --- transfer of  an input qubit state from the first to the $N$th site of a chain 
that is governed by a Hamiltonian of the form  (\ref{HamSpCh})  [or (\ref{HamHub})], at a prescribed time.  Various Hamiltonians  have been found to achieve this task \cite{QSTRevs}, one of which will be adopted later on for our purposes. In general, the input state need not be a pure state, and a measure for the fidelity of the transfer is  \cite{thesis,book}
\bea
F(\rho,\sigma) = \left ({\rm Tr}\sqrt{\sigma^{1/2}\rho\sigma^{1/2}}\right )^2,
\label{Fdef}
\eea
where $\sigma$ is the input state, and $\rho$ is the state at the output. The minimum fidelity
\bea
F_{\min} = \min_\sigma \{F(\rho,\sigma)\}, 
\eea
is obtained by taking the minimum over all the possible input qubit states. 
Using the joint concavity of the fidelity one can show that  it is sufficient to take the minimum over all the 
possible pure qubit states \cite{thesis,book} i.e.,   
\begin{subequations}
\bea
F_{\min} = \min_\psi \{F_\psi\} 
\label{Fmin}
\eea
 where 
\bea
F_\psi = \bra{\psi}\rho\ket{\psi} , 
\label{Fpsi_general}
\eea
\end{subequations}
and
\bea
\ket{\psi}  =\alpha\ket{{0}}+\beta\ket{{1}},
\label{psi}
\eea 
with $|\alpha|^2+|\beta|^2=1$ and $\alpha,\beta\in\Comp$.

Assume that the entire chain is initially prepared in the ground (vacuum) state, and the first 
spin (site) is prepared in state $\ket{\psi}$. The Hamiltonian (\ref{HamSpCh}) [or (\ref{HamHub})] preserves 
the number of spin- [or particle-] excitations, and thus we need to consider 
only the zero $\ket{\bm{0}} \equiv \prod_{j=1}^N \ket{0}_j$ and 
single excitation $\ket{\bm{j}} \equiv \hs_j^+ \ket{\bm{0}}$ 
$(\ha^{\dag}_j\ket{\bm{0}})$ sectors of the total Hilbert space.
Then the entire chain is initially in state 
$\ket{\Psi(0)} = \alpha \ket{\bm{0}} + \beta \ket{\bm{1}}$ and 
evolves in time as 
$\ket{\Psi(t)} = \hat{\mathscr U}(t)  \ket{\Psi(0)} = \alpha \ket{\bm{0}} 
+ \beta \sum_{j=1}^N A_j(t) \ket{\bm{j}}$, 
where $\hat{\mathscr U}(t)  = \exp \big[\frac{1}{i \hbar} \hat{\mathscr H}t \big]$ 
is the evolution operator, and $A_j(t) \equiv \bra{\bm{ j}}\hat{\mathscr U}(t)\ket{{\bf 1}} $.

Apparently, only the states in the single excitation sector $\{\ket{\bm{j}}\}$ evolve in time,
while the ground  (or vacuum) state $\ket{\bm{0}}$ remains unchanged. We are interested in 
the transfer of the input state from the first to the $N$th site. 
The reduced operator for the $N$th site at time $t$ is 
\bea
\rho(t) &=& {\rm Tr}_{\not {N}}[\ket{\Psi(t)}\bra{\Psi(t)}] \\ 
&=&(1-|\beta|^2|A_N|^2)\ket{0}\bra{0}+|\beta|^2|A_N|^2\ket{1}\bra{1}\nonumber\\
&&+\alpha\beta^* A_N^*\ket{0}\bra{1}+\alpha^* \beta A_N\ket{1}\bra{0},
\label{rho}
\eea
where 
\bea
A_N(t) \equiv \bra{\bm{N}}\hat{\mathscr U}(t)\ket{{\bf 1}} = \sqrt{p(t)} e^{i\varphi}, \label{An2}
\label{An}
\eea
and $p(t)$ is the probability for the excitation transfer.

Using Eqs. (\ref{Fpsi_general}), (\ref{rho}), and (\ref{psi}) we obtain
\bea
F_\psi(|\beta|^2,p,\varphi) &=& 1+|\beta|^2[-1-p+2\sqrt{p}\cos(\varphi)]\nonumber\\
&&+[2p-2\sqrt{p}\cos(\varphi)]|\beta|^4.
\label{Fpsi}
\eea
This is the fidelity in a {\em single realization} of the protocol, for a given input state $\ket{\psi}$.  Clearly, it depends on the input state (through $\beta$), the chosen QST  Hamiltonian and the presence of imperfections through $p$ and $\varphi$. 

In the absence of disorder, at the end of an ideal single realization of a QST protocol, one would have 
a well-defined amplitude $A_N=\sqrt{p_0} e^{i\varphi_0}$, with $p_0\leq 1$ (equality holds for perfect-state-transfer protocols), whereas  the phase is fixed and known, and thus, in principle,  correctable. Without loss of generality, we can assume that one compensates for $\varphi_0$ at the end of any (ideal or non-ideal) realization of the transfer, and thus 
the  fidelity (\ref{Fpsi}) can be rewritten as \cite{YBB10,Romito,Strauch,DeChiara,dapPRA10,Zwick}
\bea
F_\psi(|\beta|^2,p,\Delta\varphi) &=& 1+|\beta|^2[-1-p+2\sqrt{p}\cos(\Delta\varphi)]\nonumber\\
&&+[2p-2\sqrt{p}\cos(\Delta\varphi)]|\beta|^4,
\label{Fpsi2}
\eea
where $\Delta\varphi = \varphi-\varphi_0$. Apparently, for an ideal realization of any faithful QST protocol one has $p=p_0$ and $\Delta\varphi = 0$. 

Any physical implementation of a QST protocol will inevitably  deviate from the ideal scenario due to the presence of disorder. In principle, one can distinguish between static and dynamic disorder, which manifest themselves through the randomization of the diagonal (energies) and off-diagonal (couplings) elements of the Hamiltonian.  In the case of dynamic disorder, such a randomization varies within the time interval that the transfer takes place. On the other hand,  in the case of static disorder the randomization remains practically constant during a single realization of the transfer. The origin of such a disorder can be attributed to manufacturing errors, or to dynamic unpredictable factors (e.g., temperature fluctuations, stray fields, etc), which introduce a time-dependent disorder that does not vary appreciably over the time scale of a single transfer.  It varies, however, appreciably from realization to realization in a random way, and thus one cannot compensate for the unpredictable changes.  

The main effect of the disorder, static or dynamic, is to randomize $p$, $\varphi$  (around $p_0$ and $\varphi_0$, respectively), and thus the single-realization fidelity $F_\psi$ also becomes a random number that depends on the input state $\ket{\psi}$. For the rest of this section we will focus on static disorder.  To the best of our knowledge, so far analogous studies in the literature have relied on  the input-independent fidelity \cite{YBB10,Romito,Strauch,DeChiara,dapPRA10,Zwick}
\bea
\bar{F}(p,\Delta\varphi) = \frac{1}2+\frac{p}{6}+\frac{\sqrt{p}\cos(\Delta\varphi)}{3},
\label{FavInput}
\eea
obtained by averaging (\ref{Fpsi2}) over all the possible input states.
This average-state fidelity still varies from realization to realization and by taking an ensemble average over many realizations, the performance of a particular QST protocol is usually quantified by the ensemble averaged  input-independent quantity $\aver{\bar{F}(p,\Delta\varphi)}$. 

A faithful QST Hamiltonian should operate reliably in every single realization and irrespective 
of the input state. A question therefore  is how reliably $\bar{F}(p,\Delta\varphi)$ describes a single realization, where one typically has the transfer of a particular, and {\em a priori} unknown, qubit state $\ket{\psi}$. To answer this question, one has to compare $\bar{F}(p,\Delta\varphi)$ to the single-realization fidelity $F_\psi(|\beta|^2,p,\Delta\varphi)$ for a particular QST Hamiltonian. In the following section we present such an analysis.

\begin{figure}
\centerline{\includegraphics[scale=0.35]{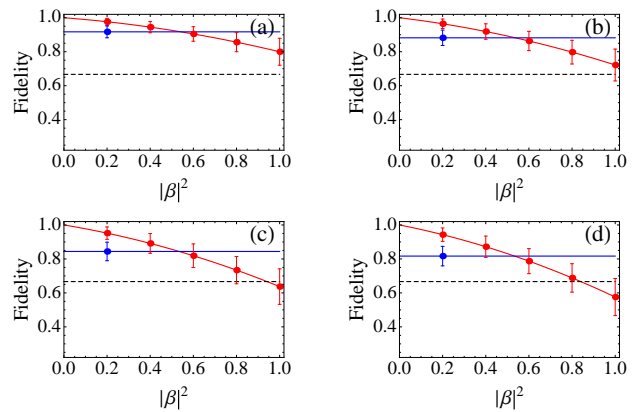}}
\caption{(Color online) The average fidelities $\langle F_\psi(|\beta|^2,p,\Delta\varphi)\rangle$ (solid curves) and  $\aver{\bar{F}(p,\Delta\varphi)}$ (horizontal solid lines) as functions of $|\beta|^2$, for different values of $N$: (a) $N=12$; (b) $N=18$; (c) $N=25$;  (d) $N=31$. The averages have been obtained on 1000 independent realizations and the error bars show the standard deviations as obtained from the corresponding distributions of $\bar{F}(p,\Delta\varphi)$ and $F_\psi(|\beta|^2,p,\Delta\varphi)$    for various values of $|\beta|^2$. The standard deviation of $\bar{F}(p,\Delta\varphi)$ is independent of $|\beta|^2$. The dashed horizontal line shows the classical threshold  $F_{cl}=2/3$. Other parameters as in Fig. \ref{distro:fig}.}
\label{beta:fig}
\end{figure}

\begin{figure}
\includegraphics[scale=0.35]{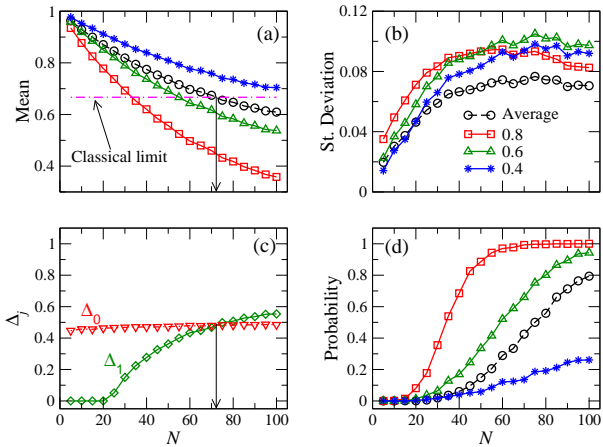}
\caption{(Color online) (a) The average fidelities $\langle F_\psi(|\beta|^2,p,\Delta\varphi)\rangle$ for various values of $|\beta|^2$ (solid curves with symbols) and  $\aver{\bar{F}(p,\Delta\varphi)}$ (dashed curve with circles) are plotted as functions of $N$. (b) As in (a) for the standard deviations. (c) The intervals $\Delta_j$ for which the average fidelity $\langle F_\psi(|\beta|^2,p,\Delta\varphi)\rangle$ is below $\aver{\bar{F}(p,\Delta\varphi)}$ (dot-dashed curve with triangles) and the classical limit (solid curve with squares), as functions of $N$. (d) The probability of failure for various values of $|\beta|^2$, as a function 
of $N$.  The vertical arrows mark the values of $N$ for which $\aver{\bar{F}(p,\Delta\varphi)}=F_{cl}$. The inset of Fig. \ref{Ndep:fig}(b) indicates the values of $|\beta|^2$ that correspond to the depicted curves.  Other parameters as in Fig. \ref{beta:fig}.}
\label{Ndep:fig}
\end{figure}

\section{Fidelity statistics}
\label{sec3}

The previous formalism applies to any QST Hamiltonian of the form  (\ref{HamSpCh}) [or (\ref{HamHub})]. 
For the sake of concreteness, however, the following analysis is focused on a well-studied  QST Hamiltonian involving \cite{QSTRevs,weNPL,Ekert04,KE05,GMN08,CVR11}
\bea
\varepsilon_k = \varepsilon,\textrm{ and } J_k=J_0\sqrt{k(N-k)},
\label{paramH}
\eea 
(see also Ref. \cite{BNT12} for the first related  proof-of-principle experiment).
It has been shown that this Hamiltonian under ideal conditions ensures the perfect transfer of a qubit state between the two ends of the 
chain, at time $\tau=\pi/(2J_0)$ i.e., 
\[
A_N(\tau) = e^{i\varepsilon\tau/\hbar}(-i)^{N-1},
\]
so that $p_0=1$ \cite{remark4}. In any physical realization of such a Hamiltonian, it is expected to exist an upper bound on the 
achievable coupling strengths $J_{\max} = \max\{J_k\}$, imposed by technological or physical constraints. Throughout our simulations we worked with the dimensionless quantities $\tilde{\varepsilon}_k\equiv  \varepsilon_k/J_{\max}$ and 
$\tilde{J}_k = J_k/J_{\max}$. We performed  a large number of realizations, each one pertaining to a sequence of independent random variables $\eta_k$ and $\xi_k$, normally distributed around zero, and with standard deviations $\sigma_\eta$ 
and $\sigma_\xi$, respectively. The energies and the couplings entering our QST Hamiltonian were randomized as follows
\bea
\tilde{\varepsilon}_k \to \tilde{\varepsilon}+\eta_k, \textrm{ and } \tilde{J}_k \to \tilde{J}_k(1+\xi_k).
\label{randH}
\eea
In each realization, we kept track of $A_N(\tau)$ obtaining thus a large sample 
of the parameters entering the fidelities. 

As mentioned above, the performance of the protocol in a single realization can be studied either in terms of the average-state  fidelity  $\bar{F}(p,\Delta\varphi)$, or the actual single-realization fidelity $F_\psi(|\beta|^2,p,\Delta\varphi)$. 
Due to the presence of disorder, both of them fluctuate from realization to realization, and in Fig.  \ref{distro:fig} we show the  corresponding distributions after 1000 realizations, for various input states (i.e., various values of $|\beta|^2$). Clearly, in many cases the distributions are rather broad, which means that the performance of the protocol in a single realization may deviate considerably from the mean fidelities $\aver{\bar{F}}$ and $\aver{F_\psi}$. Moreover, for $|\beta|^2\gtrsim 0.5$ the distributions of $F_\psi(|\beta|^2,p,\Delta\varphi)$ [see Fig. \ref{distro:fig}(a-c)] are wider than the distribution of $\bar{F}(p,\Delta\varphi)$ [see Fig. \ref{distro:fig}(f)]], and at the same time they are peaked at lower fidelities; the opposite is true for  $|\beta|^2\lesssim 0.5$. This is a typical behaviour of the distributions for fixed $N$, which suggests that $\bar{F}$  tends to overestimate (underestimate) the performance of the protocol in a single realization for states with $|\beta|^2\gtrsim 0.5$ ($|\beta|^2 \lesssim 0.5$), respectively.  

In Fig. \ref{beta:fig} we plot the ensemble averaged fidelities $\langle F_\psi(|\beta|^2,p,\Delta\varphi)\rangle$ and  $\aver{\bar{F}(p,\Delta\varphi)}$ as functions of  $|\beta|^2$, together with the classical limit $F_{cl} =2/3$ (dashed line) \cite{remark2}, for various values of $N$. In addition we plot the standard deviations of the distributions of  $F_\psi(|\beta|^2,p,\Delta\varphi)$ and  $\bar{F}(p,\Delta\varphi)$, for various $|\beta|^2$ (vertical bars).  In agreement with our observations on Fig. \ref{distro:fig} we see that for $|\beta|^2\lesssim 0.5$, $\langle F_\psi\rangle$ is above $\aver{\bar{F}}$ and decreases with increasing $|\beta|^2$,  crossing  $\aver{\bar{F}}$ at $|\beta|^2\approx 0.5$. In addition, the standard deviation of $F_\psi$ increases with increasing  $|\beta|^2$, which implies a larger dispersion of  the single realization fidelities around the mean value $\aver{F_\psi}$. As a result,  $\aver{\bar{F}(p,\Delta\varphi)}$ is not a reliable measure of the performance of the protocol under consideration, since for a broad range of input states with $|\beta|^2\gtrsim 0.5$, most likely the fidelity of the transfer in a single realization will be well below  $\aver{\bar{F}(p,\Delta\varphi)}$. More importantly, for input states with $|\beta|^2$ close to 1 and for relatively large values of $N$ (depending on the disorder under consideration), it is also highly probable that 
the fidelity of the transfer in a single realization of the given QST protocol will be below the classical limit $F_{cl}$, which implies that the particular state transfer has failed, yet $\aver{\bar{F}}$ is well above $F_{cl}$ indicating faithful QST [e.g., see Figs.  \ref{beta:fig}(c) and (d)]. 
 
 \begin{figure}
\centerline{ \includegraphics[scale=0.6]{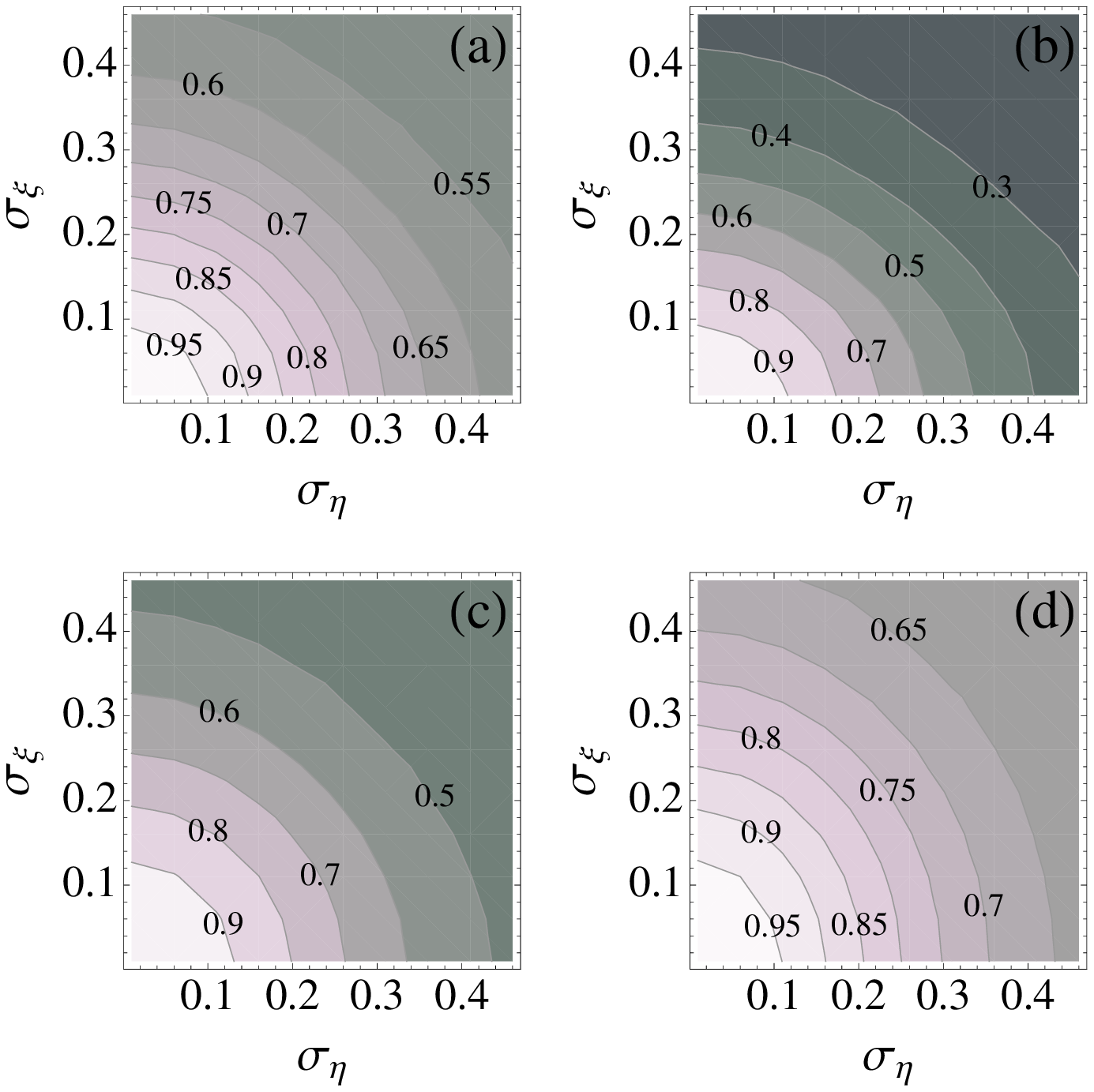}}
 \centerline{\includegraphics[scale=0.35]{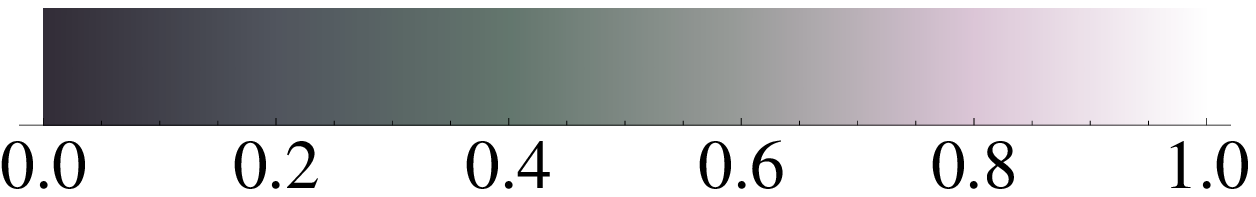}}
\caption{(Color online) The average fidelities $\aver{\bar{F}(p,\Delta\varphi)}$ (a) and $\langle F_\psi(|\beta|^2,p,\Delta\varphi)\rangle$ for $|\beta|^2=0.8$ (b), 
$|\beta|^2=0.6$ (c) and $|\beta|^2=0.4$ (d), as functions of disorder for $N=15$. The averages have been obtained on 1000 independent realizations.}
\label{MeanDis:fig}
\end{figure}
 
\begin{figure}
 \includegraphics[scale=0.6]{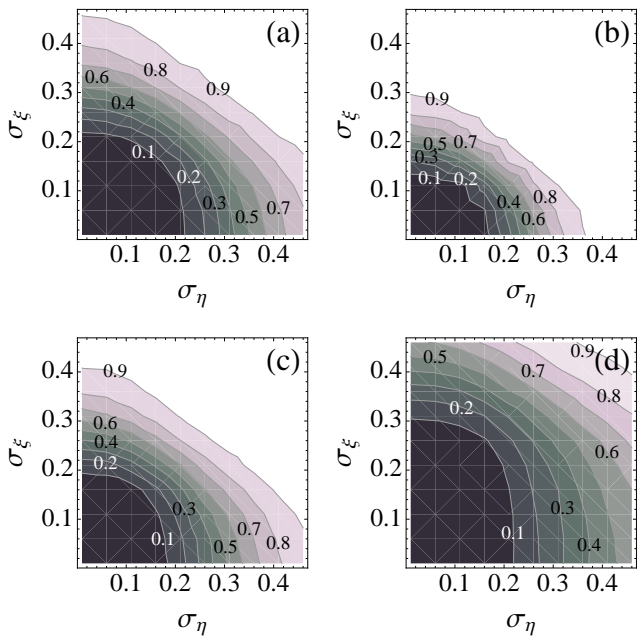}
\caption{(Color online) The probability of failure as estimated by $\aver{\bar{F}(p,\Delta\varphi)}$ (a), and $\langle F_\psi(|\beta|^2,p,\Delta\varphi)\rangle$ for $|\beta|^2=0.8$ (b), $|\beta|^2=0.6$ (c) and $|\beta|^2=0.4$ (d), as a function of disorder for $N=15$. The averages have been obtained on 1000 independent realizations. Successive contours differ by 0.1 and the color scale is as in Fig. \ref{MeanDis:fig}.}
\label{PFDis:fig}
\end{figure}

\begin{figure}
 \includegraphics[scale=0.6]{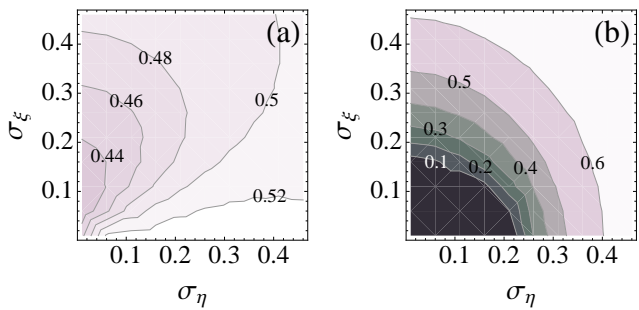}
\caption{(Color online) The quantities $\Delta_0$ (a) and $\Delta_1$ (b) as functions of disorder for $N=15$. The averages have been obtained on 1000 independent realizations.}
\label{CritDis:fig}
\end{figure}

To make this point clearer, in Fig. \ref{Ndep:fig}(a) we plot the dependence of  $\langle F_\psi(|\beta|^2,p,\Delta\varphi)\rangle$ and  $\aver{\bar{F}(p,\Delta\varphi)}$ on the 
number of sites in the chain, for a fixed disorder. In all cases, the ensemble averaged fidelities decrease with $N$, but  for input states with 
$|\beta|^2\gtrsim 0.5$,  $\langle F_\psi\rangle$ decreases considerably faster than $\aver{\bar{F}}$, crossing the classical limit at smaller values of $N$. 
Moreover, the corresponding standard deviations increase with $N$ [see Fig. \ref{Ndep:fig} (b)], and in general the ones of  $F_\psi$ for $|\beta|^2\gtrsim 0.5$ are larger than the standard deviation of $\bar{F}$, for all the values of $N$. 

As depicted in Fig. \ref{beta:fig}, for any fixed $N$, $\langle F_\psi(|\beta|^2,p,\Delta\varphi)\rangle=1$ at $|\beta|^2=0$, and decreases with increasing $|\beta|^2$, whereas $\aver{\bar{F}(p,\Delta\varphi)}$ and $F_{cl}$ are independent of $|\beta|^2$. 
The values of $|\beta|^2\in[0,1]$ at which $\langle F_\psi(|\beta|^2,p,\Delta\varphi)\rangle$ crosses $\aver{\bar{F}(p,\Delta\varphi)}$ and $F_{cl}$, are of particular interest since they essentially characterize the class of input qubit states for which $\langle F_\psi(|\beta|^2,p,\Delta\varphi)\rangle$ becomes smaller than the other two fidelities.  For such a characterization, we can introduce here the quantity
\bea
\Delta_j = \left\{
\begin{array}{rl}  
1-B_j^{(c)}, & \text{if }   0\leq B_j^{(c)}\leq 1\\
0, & \text{otherwise} 
\end{array} \right. ,
\eea
where $B_j^{(c)}$, with $j\in\{0,1\}$, are the roots of the polynomials $S_0(|\beta|^2)=\langle F_\psi(|\beta|^2,p,\Delta\varphi)\rangle-\aver{\bar{F}(p,\Delta\varphi)}$ and $S_1(|\beta|^2)=\langle F_\psi(|\beta|^2,p,\Delta\varphi)\rangle-F_{cl}$, with respect to $|\beta|^2$. For all the tested parameters and noises in our simulations, these polynomials had at most one root in the interval $[0,1]$, and thus $\Delta_j$ were uniquely defined. As shown in Fig. \ref{Ndep:fig}(c), $\Delta_0$ does not vary considerably with $N$, and throughout our simulations it was around $0.5$. This means that for all $N$, there is always a class of  input qubit states with $|\beta|^2>B_0^{(c)}$, where $B_0^{(c)}\approx 0.5$,  for which $\langle \bar{F}(p,\Delta\varphi)\rangle$ overestimates the performance of the protocol relative to $\langle F_\psi(|\beta|^2,p,\Delta\varphi)\rangle$.  Moreover, for relatively small values of $N$, there is no root for the polynomial $S_1$ in $[0,1]$, and thus $\Delta_1=0$,  which means essentially that $\langle F_\psi(|\beta|^2,p,\Delta\varphi)\rangle\geq F_{cl}$, for all $|\beta|^2\in[0,1]$. As we increase $N$, however,  the polynomial $S_1$  acquires a root $B_1^{(c)}\in [0,1]$, which appears close to 1 for $N\approx 23$, and moves rapidly toward 0.4, with increasing $N$. As a result, the range of values of $\beta$ for which $\langle F_\psi(|\beta|^2,p,\Delta\varphi)\rangle < F_{cl}$ is getting wider with increasing $N\gtrsim 23$, whereas at the same time, one has $\langle \bar{F}(p,\Delta\varphi)\rangle < F_{cl}$ only for $N\gtrsim 73$. 
Hence, for $23\lesssim N\lesssim 73$ the fidelity  $\langle \bar{F}(p,\Delta\varphi)\rangle$ predicts incorrectly the transfer of input qubit states with $|\beta|^2\gtrsim B_1^{(c)}$, whereas the transfer of such qubit states under the particular QST protocol is most likely impossible (i.e., the fidelity in a single realization is below the classical limit). 

Of course, as mentioned before, there are statistical deviations from realization to realization and thus in our simulations we have also kept track of the number of realizations for which  
$F_\psi(|\beta|^2,p,\Delta\varphi) \geq F_{cl}$ and  $\bar{F}(p,\Delta\varphi) \geq F_{cl}$. Dividing by the total number 
of realizations, we thus have an estimate of the probabilities of failure for the protocol under consideration and for various input states. As shown in Fig. \ref{Ndep:fig}(d), for all the states with $|\beta|^2 > B_0^{(c)}$ (see the curves with squares and triangles corresponding to $|\beta|^2=0.8\textrm{ and }0.6$, respectively), the probability of failure that is estimated according to $F_\psi(|\beta|^2,p,\Delta\varphi)$ can be considerably higher than the one that is obtained from $\bar{F}(p,\Delta\varphi)$ (dashed curve with circles). The fidelity 
$\bar{F}(p,\Delta\varphi)$ predicts reliably the probability of failure only in two cases. Namely, either for states with $|\beta|^2 \leq B_0^{(c)}$ for all 
$N$ (see curve with stars corresponding to $|\beta|^2 = 0.4$), or for all the input qubit states when $N\lesssim 10$.  

\begin{figure*}
\includegraphics[scale=0.6]{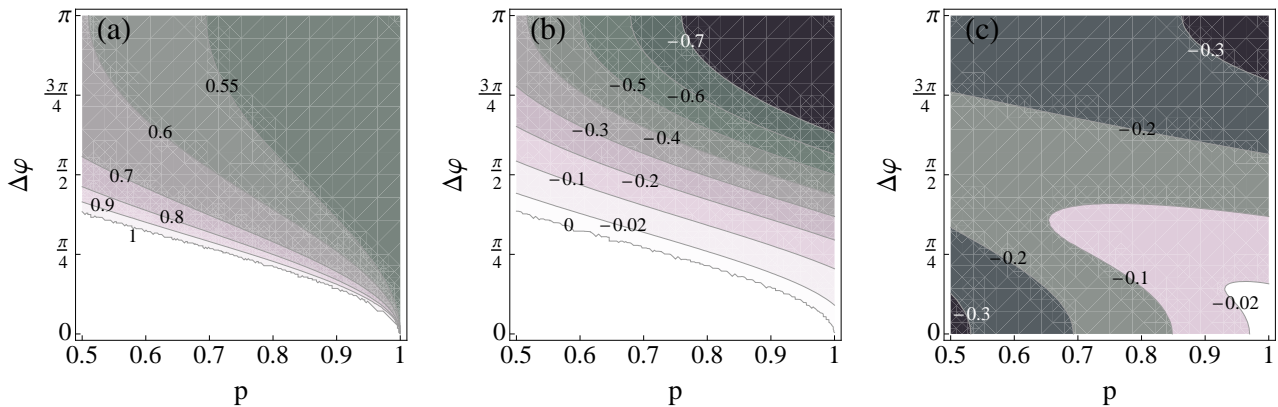}
\caption{(Color online) The values of $|\beta|^2$ that minimize $F_\psi(|\beta|^2,p,\Delta\varphi)$ (a)  as well as  the fidelity differences $F_{\min}(p,\Delta\phi)-F(1, p, \Delta\phi)$ (b)  and $F_{\min}(p,\Delta\phi)-\bar{F}(p,\Delta\phi) $ (c),  for various combinations of $p$ and $\Delta\phi$. The plots are mirror-symmetric with respect to $\Delta\varphi=0$, and can be extended to $\Delta\varphi<0$ by the transformation $\Delta\varphi\to-\Delta\varphi$.}
\label{bf:fig}
\end{figure*}

Our discussion so far has been restricted to a particular combination of diagonal and off-diagonal static disorder (i.e., fixed $\sigma_\eta$ and $\sigma_\xi$). 
In Fig. \ref{MeanDis:fig}(a), we present a contour plot of  $\aver{\bar{F}(p,\Delta\varphi)}$ as a function of $\sigma_\eta$ and $\sigma_\xi$, 
whereas Figs. \ref{MeanDis:fig}(b-d) are the corresponding plots of $\aver{F_\psi(|\beta|^2,p,\Delta\varphi)}$, for various values of $|\beta|^2$.  In all cases, the  fidelities drop with increasing disorder, but for $|\beta|^2=0.6$ and $|\beta|^2=0.8$ $\aver{F_\psi(|\beta|^2,p,\Delta\varphi)}$ decreases considerably faster than $\aver{\bar{F}(p,\Delta\varphi)}$, whereas the opposite is true for $|\beta|^2=0.4$.  Analogous contour plots for the probability of failure are depicted in Fig. \ref{PFDis:fig}. Again we see that as we increase the disorder for $|\beta|^2=0.6$ and $|\beta|^2=0.8$,  the probability of failure for $\aver{F_\psi(|\beta|^2,p,\Delta\varphi)}$ increases considerably faster than the probability of failure for $\aver{\bar{F}(p,\Delta\varphi)}$. The latter seems to overestimate the  probability of failure only when $|\beta|^2\lesssim 0.5$ for the specific value of $N$.  Finally,  Fig. \ref{CritDis:fig}   shows that the interval of values for $|\beta|^2$ where $\aver{F_\psi(|\beta|^2,p,\Delta\varphi)}<\aver{\bar{F}(p,\Delta\varphi)}$ does not vary appreciably with the disorder (it is around 0.5). On the contrary, the corresponding interval where $\aver{F_\psi(|\beta|^2,p,\Delta\varphi)}<F_{cl}$ becomes wider with increasing disorder. 

To summarize, all of the above observations confirm  the fact that the average-state fidelity $\bar{F}(p,\varphi)$, usually employed in studies on QST,  is not a reliable measure for the performance of QST protocols in a single realization for all the possible levels of disorder, and for all possible input qubit states. It tends to overestimate the performance of QST protocols for a broad class of qubit states, and it may even lead to faulty conclusions with respect to the classical limit.  As will be discussed 
in the following section, a more reliable measure for the performance of QST protocols is provided by $F_{\min}$, or even by the probability $p$ in certain cases.

\section{Reliable measures of state transfer}
\label{sec4}

The most reliable measure for the quantification of  QST is the minimum fidelity (\ref{Fmin}). 
The minimization of the single-realization fidelity $F_\psi$, over all the possible input qubit states is straightforward in the 
absence of disorder where $p=p_0$ and $\Delta\varphi=0$. Indeed, one can readily show that $F_\psi(|\beta|^2,p_0,0)$ is a monotonically decreasing function of $|\beta|^2\in[0,1]$, which attains its minimum at $|\beta|^2=1$ and is equal to $p_0$. This means that in the case of ideal realizations, the transfer of the excitation (probability) implies transfer of the quantum state with precisely the same fidelity. In the case of perfect QST Hamiltonians,  $p_0=1$ for all input states. 

In general, the transfer of the excitation is quantified by the probability $p$, which in the presence of disorder static or dynamic,  is  a random variable in the interval $[0,p_0]$, whereas $\Delta\varphi$ is a random variable in $[-\pi,\pi]$. In this case, the transfer of the excitation does not necessarily  imply transfer of the state (which  in addition to bit information also carries phase information) \cite{remark3}.
For relatively weak disorder, the distributions of $p$ and $\Delta\varphi$ are expected to be peaked at the corresponding ideal values $p_0$ and  $0$, respectively, and how fast the distributions drop depends on the type of noise under consideration. From the mathematical point of view, irrespective of the Hamiltonian and the disorder under consideration, $F_\psi(|\beta|^2,p,\Delta\varphi)$ is a stochastic function, which for any given $p\in[0,1]$ and $\Delta\varphi\in[-\pi,\pi]$ attains its maximum for $\beta=0$. Moreover, its first derivative with respect to $|\beta|^2$ vanishes for $|\beta|^2=B^*\in \Real$, where
\bea
B^* = \frac{1+p-2\sqrt{p}\cos(\Delta\varphi)}{4[p-\sqrt{p}\cos(\Delta\varphi)]}, 
\eea
while it is strictly negative for $|\beta|^2<B^*$, and strictly positive for 
$|\beta|^2>B^*$. Thus the fidelity $F_\psi$ exhibits its minimum  at $|\beta|^2=B^*$ if $B^*\in [0,1]$ 
and at $|\beta|^2=1$ otherwise. The corresponding map  for various combinations of $p$ and $\Delta\varphi$ is shown in 
Fig. \ref{bf:fig}(a). Clearly, for a broad range of values for $p$ and $\Delta\varphi$, the fidelity is minimized for $|\beta|^2=1$, but it does also exist a non-negligible range of values for which the fidelity is minimized for input qubit states  with  
$|\beta|^2< 1$. In the former case, the corresponding minimum value is $F_{\min}(p,\Delta\varphi)=p$  whereas in the latter 
\bea
F_{\min}(p,\Delta\varphi) = \frac{1}{2}+\frac{\sqrt{p}}{2}\cos(\Delta\varphi)-\frac{(1-p)^2}{8[p-\sqrt{p}\cos(\Delta\varphi)]}.\nonumber\\
\eea

Let us now compare $F_{\min}(p,\Delta\varphi)$ to $F_\psi(1, p, \Delta\varphi) = p$, which corresponds to the transfer of probability (excitation), as well as to $\bar{F}(p,\Delta\varphi) $. 
The differences are shown in Fig. \ref{bf:fig}(b) and Fig. \ref{bf:fig}(c), 
respectively \cite{remark1}. Clearly, when the  disorder is such that the random phase $\Delta\varphi$ is highly peaked around 0, we have $F_{\min}(p,\Delta\phi)=p$,  for all values of $p\in[0.5,1]$. Hence, in this case one can safely consider the transfer of probability as a measure for the transfer of the state.  For any other combination of $\Delta\phi$ and $p$, we have  $F_{\min}(p,\Delta\phi)<p$, and thus $F_{\min}(p,\Delta\varphi)$ has to be 
used for the quantification of the QST in a single realization. By contrast, according to Fig. \ref{bf:fig}(c), there is only a very narrow regime of combinations for $p$ and $\Delta\varphi$, where $ |F_{\min}(p,\Delta\phi)-\bar{F}(p, \Delta\phi)|\leq 0.02$.  This regime is actually very close to the ideal scenario, and is hard to be achieved in practice.

\begin{figure}
\includegraphics[scale=0.5]{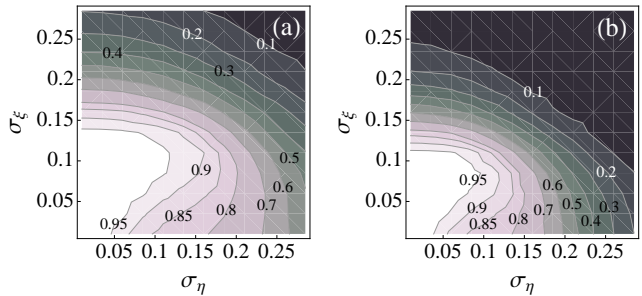}
\caption{(Color online) Density plot of probability (\ref{probPF}) as a function of the diagonal ($\sigma_\eta$) and off-diagonal ($\sigma_\xi$) disorder, for a chain of length  (a) $N=12$ and (b) $N=21$.  The probabilities have been estimated for $\epsilon = 10^{-2}$ on  $1000$ realizations.}
\label{PvsFmin:fig}
\end{figure}

It is worth emphasizing here that the main observations of this section so far apply to any QST protocol  
for which the single-realization fidelity is given by Eq. (\ref{Fpsi}), irrespective of the type of noise. In practice, for a certain QST Hamiltonian 
the parameters $p$ and $\Delta\varphi$ vary from realization to realization, with $p$ being always an upper bound on $F_{\min}$. 
Hence, to be on the safe side, one has always to quantify the performance of a QST Hamiltonian in terms of $F_{\min}$. 
The above results, however, suggest that  it does exist a regime of diagonal and off-diagonal disorders, where  $F_{\min}$ 
seems to be well approximated by $p$. Of course, the details of this regime depend on the QST Hamiltonian under 
consideration as well as on the modelling of  the disorder. For instance, we have estimated numerically for the model of 
Sec. \ref{sec3}  the probability \cite{remark1}
\bea
\textrm{Prob}[p>1/2 \cap |p - F_{\min}(p,\Delta\phi)| < \epsilon]
\label{probPF}
\eea
for $\epsilon\ll 1$,
as a function of $\sigma_\eta$ and $\sigma_\xi$ for two different values of $N$. As shown  in Fig. \ref{PvsFmin:fig},  in both cases the probability (\ref{probPF}) exceeds $90\%$ for weak disorder, which means that in this case the transfer of the excitation (probability) can describe reliably the transfer of the state up to error $\epsilon$. By contrast, the  probability $\textrm{Prob}[p>1/2\, \cap\, |\bar{F} - F_{\min}| < 10^{-2}]$, is practically negligible and is not shown here. Once more therefore we see that   
$\bar{F}(p, \Delta\phi)$ is not a reliable measure for the QST under realistic conditions, since it tends to overestimate the performance of the protocol, relative to $F_{\min}(p,\Delta\phi)$.

\section{Conclusions}
\label{sec5}

We have presented a statistical analysis on the performance of QST Hamiltonians in a single realization and in the presence of 
static disorder.  It has been shown that the average-state fidelity, usually employed in the literature for 
related studies, may fail to describe accurately the performance of QST Hamiltonians in a single realization. Most importantly, for a large class of input states it may also lead to faulty conclusions about the success of the transfer. For the sake of concreteness our analysis has been restricted to a particular well-studied QST Hamiltonian, and to a certain model of disorder. This allowed us to quantify the deviations of the average-state fidelity from the actual performance of the protocol in single realizations, as well as to identify the class of input qubit states and the size of the chains for which these deviations become fatal.

Analogous  conclusions are expected to be valid  for other Hamiltonians and other models of disorder as well --- albeit perhaps with some quantitative differences ---  and suggest that any studies on the robustness of QST Hamiltonians in the presence of noise have to rely on the minimum fidelity, while for weak disorder the transfer of the excitation (or probability) may  be a rather reliable measure as well. In view of the present results, some of the previous studies on the classification of various QST Hamiltonians based on the average-state fidelity, may have to be revisited.


\begin{thebibliography}{99}
\bibitem{remark4}
Strictly speaking, for the QST Hamiltonian under consideration one does not need the aforementioned assumption for the chain to be 
initially prepared in the ground (vacuum) state. State transfer is expected to take place irrespective of the initial state of the 
chain, provided that the input spin (site) is initially decorrelated from the rest of the chain \cite{design} .  The main 
reason is that the evolution operator at time $\tau$ reduces to a permutation (up perhaps to an unimportant global phase).
\bibitem{remark2}
The classical limit of 2/3 is the best that in principle can be achieved  by a direct projective measurement on the qubit state on one end of the chain, and transfer of the outcome via classical communication to the other end, where the qubit can be prepared according to the outcome.  The limit of $1/2$  corresponds to a random guess of the qubit state $\ket{0}$ or $\ket{1}$, with no measurements \cite{MP95}.
\bibitem{remark3} For instance, even in the case of  perfect transfer of the excitation (i.e., for $p=1$), 
Eq. (\ref{FavInput}) yields $\aver{\bar{F}} = 2/3$ when the phase is completely randomized i.e., for  $\aver{\cos(\Delta\varphi)} =0$.
\bibitem{remark1}
We are focusing on cases where the transfer of probability exceeds  0.5 that corresponds to random guess. See also remark \cite{remark2}.
\end{thebibliography}
\end{document}